# Laser-induced electron diffraction in the over-the-barrier ionization (OBI) regime

B. Belsa[1,\*], K.M. Ziems[2,3,4,\*], A. Sanchez[1], K. Chirvi[1], X. Liu[1], S. Gräfe[2,3], J. Biegert[1,5,†]

[1]*ICFO - Institut de Ciencies Fotoniques, The Barcelona Institute of Science and Technology, 08860 Castelldefels (Barcelona), Spain.*
[2]*Institute of Physical Chemistry, Friedrich-Schiller-Universität Jena, 07743 Jena, Germany.*
[3]*Abbe Center of Photonics, Friedrich-Schiller-Universität Jena, 07743 Jena, Germany.*
[4]*Max Planck School of Photonics, 07743 Jena, Germany.*
[5]*ICREA, Pg. Lluís Companys 23, 08010 Barcelona, Spain.*

[\*]*These authors contributed equally to this work.*

[†]To whom correspondence should be addressed to. Email: jens.biegert@icfo.eu

## Abstract

Large polyatomic molecules typically exhibit low ionization potentials, $I_p$, leading to over-the-barrier ionization (OBI) already at relatively low intensities (~$10^{13}$ W/cm$^2$). We revisit laser-induced electron diffraction (LIED) in the over-the-barrier ionization (OBI) regime and answer the question of whether imaging of molecular structure is still possible with LIED. We employ a hydrogen-like model system mimicking a molecule with low $I_p$ using a *classical trajectory-based model* (CT) that incorporates the Coulomb potential; we also use the numerical solution to the *time-dependent Schrodinger equation* (TDSE). Specifically, we adopt the Fourier transform variant of LIED (FT-LIED) to show that even a significant contribution of short trajectories in the OBI regime does not preclude structure retrieval from strong-field diffractive patterns. This theoretical investigation shows that LIED can be well described by the *classical recollision model* even when ionization occurs within the OBI regime. This study paves the way towards strong-field imaging of chemical transformations of large polyatomic molecules in real-time based on strong-field electron recollision.

## I. Introduction

Strong-field physics (SFP) provides enticing opportunities to image molecular structure [1–3] with picometer spatial and femto- to attosecond temporal resolution. A prominent example of such a successful method is laser-induced electron diffraction (LIED). Information about the molecule's structure is largely inferred by invoking the classical three-step recollision model to map measured electron momenta to time and space. This implies electron recollision occurs under quasi-static (tunneling) conditions. Application of strong-field molecular imaging methods to larger and more complex molecules is faced with the challenge that large field strengths need to be used for recollision (imaging) to occur, but the ionization potential of molecules decreases with size. This places the measurements away from the quasi-static (tunneling) regime into the over-the-barrier regime, which means that the initial velocities of electrons cannot be assumed to be zero anymore. The question then is whether the molecular structure can still be extracted and if the conditions lead to larger bond errors.

The regime of strong-field ionization is commonly classified by the Keldysh parameter, $\gamma = \sqrt{I_\mathrm{p}/(2U_\mathrm{p})}$ [4]. Here, $I_\mathrm{p}$ is the ionization potential of the atom or molecule and, $U_\mathrm{p} = E_0^2/(2\omega)^2$, is the ponderomotive energy (i.e., the average kinetic energy of a free electron in an oscillating electric field, with an angular frequency of $\omega$ and $E_0$ being the electric field strength). This metric distinguishes the multi-photon ionization (MPI; $\gamma \gg 1$) from the tunneling ionization (TI; $\gamma \ll 1$) regime. H. Reiss introduced additional unit-less parameters to define more precisely the upper and lower bounds on the applicability of the simple Keldysh metric [4]. Of particular interest are the amplitude parallel to the propagation direction of the field, $\beta_0 = U_\mathrm{p}/(2\omega c)$, to differentiate the regime of non-dipole effects (when $\beta_0 \geq 1$) and the intensity parameter, $z_f = 2U_\mathrm{p}/(mc^2)$, indicating the non-relativistic domain (when $z_f \geq 1$ ) [5].

When operating deep in the quasi-static, or tunnel ionization (TI), regime, several important SFP phenomena occur that are (reasonably) well-described by the *classical recollision model* or *simple-man's model (SM)* (see [1,6–12] and references therein). In this well-known model, a photoelectron is born via tunnelling through the lowered quasi-static barrier with zero kinetic energy. The laser field strength dominates the action of the electron; i.e., the Coulomb field of the ion is entirely ignored. Depending on the time of emission relative to the laser field cycle, this results in electrons that escape directly, or electrons which return to the ion [8]. In the latter case, the electron may: (i) recombine with the ion and emit a high energy photon, leading to the generation of high-order harmonics of the laser radiation [12]; (ii) inelastically scatter off the ion and cause subsequent, additional excitation or ionization of the parent cation through non-sequential double ionization (NSDI) [13]; or (iii) elastically scatter off the parent ion leading to an additional momentum transfer, $q$, between the electron and the parent ion. The latter phenomenon is a process known as laser-induced electron diffraction (LIED) [14,15]. The *Simple Man's* model allows classical trajectories to be mapped to experimental features and, ultimately, extract electronically and/or geometrical information of the target.

This work focuses on (iii), and in particular on the Fourier transform variant of the LIED process, since the target's elastically back-scattered electrons are utilized to self-image molecular structure. This scattering process is coherent and in contrast to the incoherent scattering by an electron beam, like in gas-phase electron diffraction [16]. To penetrate past the valence electron cloud, also called hard collisions, and, subsequently, pinpoint the atomic cores in the molecular structure, we need the LIED electron to *return* at backscattered angles ($\theta_r$ = 45 – 180°) and with sufficient impact energies [17]. Such impact energies were demonstrated with mid-infrared (MIR) driving lasers ($\lambda \geq 3$ μm) at high intensities (~$10^{14}$ W/cm$^2$) [18] by taking advantage of the *return* energy scaling linearly with intensity but quadratically with wavelength, $E_r \propto U_p \propto I\lambda^2$. Such scaling is however severely limited by the ionization threshold $I_p$ of the target. At larger intensity, the tunneling barrier is completely suppressed, and over the barrier ionization (OBI) takes place. This causes the electron wave function to escape quasi-barrierless [19]. In Figure 1, we sketch the OBI regime. The barrier-suppression ionization (BSI) field strength for a δ potential is $E_{BSI} = I_p^2/(4Z)$. We note that for the Coulomb potential, OBI occurs at even lower field strength. Thus, for large molecules, i.e. for low $I_p$ targets, already relatively small field strengths cause a transition from TI to OBI. For example, naphthalene, with an $I_p$ of 8 eV, reaches the OBI regime already at $E_{BSI}$ = 0.02 a.u. (i.e., 1.6×$10^{13}$ W/cm$^2$); at the same time, for a driving laser of 3.2 μm, this translates in a $U_p$ of only 14 eV. Such electron impact energies do not achieve sufficient momentum transfer for the geometric resolution of the target structure. Thus, imaging large-polyatomic molecules with recollision imaging methods like LIED requires understanding whether image reconstruction is possible in the OBI regime. From a practical point of view, this raises the question of whether the Simple Man's model, which is commonly applied to retrieve structural information, can be evoked in the OBI regime and what errors this incurs. Especial care has to be taken regarding the conversion of *laboratory* to *laser polarization frame* as during the rescattering the field is additionally imprinted on the electron momenta. Hence, measured momenta include two contributions: (i) a momentum shift caused by the electron scattering off the target molecule and (ii) a momentum shift caused by the laser's vector potential at the time of rescattering. The latter fluctuates during the laser cycle, imparting different momentum at different times of rescattering during the laser cycle.

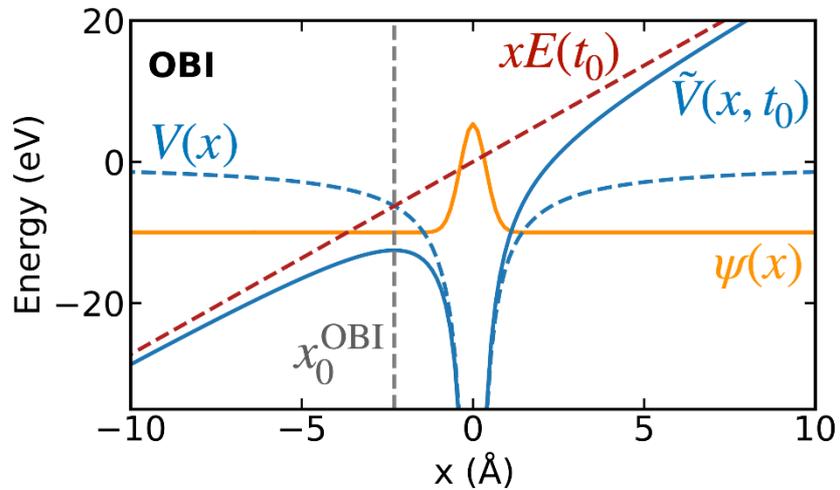

**Figure 1. Over-the-barrier ionization (OBI).** The electron-laser field interaction ($xE(t_0)$) (red dashed line) distorts the Coulomb potential ($V(x)$, blue dashed line), resulting in a dressed Coulomb potential (here shown for peak laser strength $t = t_0$, $\tilde{V}(x,t_0) = V(x) +$

$xE(t_0)$, solid blue line). When the distortion of the Coulomb potential is such deformed that tunneling barrier is completely suppressed (as shown here), the electron wave function ($\psi(x)$ yellow line) is able to escape quasi-freely, giving place to the so-called OBI regime. For more details, see text.

In the present work, we answer these questions with a theoretical investigation of LIED for low $I_p$ targets in the OBI regime. For this, we employ a hydrogen-like atom with $I_p$ = 10 eV in MIR linearly polarized laser fields. We compare the results of the (i) *classical recollision model,* or *Simple Man's model* (SM), which neglects the Coulomb potential with (ii) a *classical trajectory model* (CT) incorporating the Coulomb potential explicitly together with the full (iii) quantum dynamics from the numerical solution of the *time-dependent Schrödinger equation* (TDSE). Finally, we use *Fourier transform* LIED (FT-LIED) [18,20–22] to examine the influence of short trajectories in the OBI regime on the structure retrieval in LIED. In particular, we focus on the effect of the driving wavelength on the LIED process and structure retrieval. In the following, we will show that the SM can be used for LIED even in the OBI regime without significant error in the structure analysis. This finding enables a development of LIED for large and complex polyatomic molecules with low $I_p$.

The structure of the paper is the following: In section II, we describe the computational details and methods. The results are discussed in section III, starting with the a) trajectory-based methods, followed by the b) TDSE results and c) the FT-LIED model that accounts for short trajectories in the LIED analysis. The last section presents an error analysis of the SM-based LIED structure retrieval method. Finally, the conclusion is given in section IV.

## II. Computational Details and Methods

In the following, we introduce the three different levels of theory employed to analyze the properties of returning electrons in the OBI regime: (i) the *classical recollision model*, *or Simple Man's model* (SM), omitting the contribution of the Coulomb potential completely (ii) a *classical trajectory model* (CT), and (iii) quantum dynamics obtained from the direct numerical solution of the *time-dependent Schrödinger equation* (TDSE). In the following, all equations are in atomic units unless otherwise stated.

In the CT and TDSE, we employ a screened one-dimensional atomic model system with the potential (see Figure 1):

$$V(x) = -\frac{Z}{\sqrt{x^2 + \alpha}}, \tag{1}$$

being $Z$ the nuclear charge and $\alpha$ the smoothing parameter. The results presented here are for a hydrogen-like atom with $Z = 1$ and $\alpha = 4.08482$ a.u. resulting in an ionization potential of $I_p$ =

10 eV as common range for many molecules. The dressed Coulomb potential, $\tilde{V}(x,t)$, is defined as (see Figure 1):

$$\tilde{V}(x,t) = V(x) + xE(t) \tag{2}$$

The field strength threshold for the TI to OBI transition is $E_{\text{BSI}} = I_p^2/(4Z) = 0.034$ a.u. ($I_{\text{BSI}} = 4\times10^{13}$ W/cm²).

### a. Trajectory-based methods

For the trajectory-based methods, SM and CT, with electric field

$$E(t) = E_0 \cos(\omega t), \tag{3}$$

an electric field strength of $E_0 = 0.053$ a.u., ($I = 1\times10^{14}$ W/cm²), ionization occurs in the OBI regime for our setup. We aim at evaluating the effect of the wavelength scaling, thus, we perform calculations for different optical frequencies $\omega = $ 0.023, 0.014, 0.011 a.u. (i.e., $\lambda = $ 2.0, 3.2 and 4.0 µm). The values were chosen to mimic common experimental setups [1,17,23–30]. The parameters are summarized in Table 1, together with their corresponding ponderomotive energy $U_p = E_0^2/(2\omega)^2$. We note that non-dipole effects are noticeable already at $\lambda = 4.0$ µm for the relevant field strengths; this will be discussed later in the text.

Table 1. The Keldysh parameter ($\gamma$) and parameters indicating strong-field response ($z_f$ and $\beta_0$, see H. Reiss [4]) together with the $U_p$ value for the corresponding field strength in the OBI regime, $E_{\text{OBI}} = 0.053$ a.u., for the different wavelengths ($\lambda$) employed in the present work.

| $\lambda$ (µm) | $\gamma$ | $z_f$ | $\beta_0$ | $U_p$ (eV) |
|---|---|---|---|---|
| **2.0** | 0.37 | $1.5\times10^{-4}$ | 0.22 | 37 |
| **3.2** | 0.23 | $3.7\times10^{-4}$ | 0.89 | 94 |
| **4.0** | 0.18 | $5.8\times10^{-4}$ | 1.76 | 149 |

Within a time interval of $t_0 \in [-0.25, 0.25]$ opt. cycles, (the positive half-cycle of the electric field) 20000 classical trajectories are started.

The dynamics of the released electron trajectories are determined by Newton's equation of motion, $p = \frac{\partial x}{\partial t}$. The acceleration $\frac{\partial p}{\partial t} = F(x,t)$ is obtained in the as follows: In SM, only the force exerted by the electric field, $\partial p/\partial t = -E(t)$, is considered, whereas in the CT, the Coulomb potential is explicitly included, thus $\partial p/\partial t = -\nabla_x \tilde{V}(x,t)$.

The initial conditions at time of ionization for an individual trajectory, $t_i$, are as follows: Within the SM, the initial conditions of each trajectory are simplified as $x_0(t_i) = 0$, $p_0(t_i) = 0$. In the CT, the initial conditions differ: The initial spatial position for OBI is approximated by the local maximum of the dressed Coulomb potential, $\nabla_x \tilde{V}(x, t_i) = 0$ (on top of the quasi-static barrier formed by the field):

$$x_0^{OBI}(t_i) = -\sqrt{\frac{Z}{E(t_i)}} \quad (4)$$

for a positive half-cycle of the electric field (see Figure 1). The initial OBI velocity, $p_0^{OBI}(t_i)$, for a positive half-cycle is determined by the energy difference (excess energy) of the dressed potential at the OBI exit, $\tilde{V}(x_0^{OBI}(t_i), t_i)$, and the ionization potential, $I_p$ giving

$$p_0^{OBI}(t_i) = -\sqrt{2\left(-I_p - \tilde{V}(x_0^{OBI}(t_i), t_i)\right)}. \quad (5)$$

A fourth-order Runge-Kutta algorithm is utilized to solve the respective Newton's equations in SM and CT for two full optical cycles using a time step of $\Delta t = 0.27$ as. Each trajectory is weighted by the ionization rate according to Ammosov, Delone and Krainov (ADK) [31]:

$$\Gamma \propto e^{-\frac{4\sqrt{2}\,I_p^{3/2}}{3E(t)}}, \quad (6)$$

approximating the ionization probability. A comparison with the empirical formula for TI and OBI regimes and described by Tong and Lin [32] showed no significant difference in the results discussed here.

The calculated trajectories are classified as *returning trajectories* once they return to the nucleus by crossing $x_{ret} = 0$. They are subsequently analyzed regarding their *return* energy, $E_{ret}$, and *return* time, $t_{ret}$. Additionally, the *rescattering* energy, $E_{resc}$, can be calculated by incorporating the vector field at the time of *return*, $A(t_{ret})$:

$$E_{resc} = \frac{1}{2}\left(p(t_{ret}) + A(t_{ret})\right)^2. \quad (7)$$

In the CT, due to the exact incorporation of the Coulomb potential, the momentum of *rescattering* trajectories shows a *funnel effect* upon return to the nucleus: the electron is accelerated in and out of the Coulomb potential leading to a cusp in the momentum. The $E_{ret}$ is, thus, extracted by fitting a trigonometric function to the relevant optical cycle that describes the *returning* electron. More details are given in Appendix A.

### b. *Numerical solution of the time-dependent Schrödinger equation (TDSE)*

Besides the classical models described above, we also investigate the quantum dynamics as obtained by a numerical solution of the time-dependent Schrödinger equation (TDSE)

$$i\frac{\partial}{\partial t}|\Psi(t)\rangle = H(t)|\Psi(t)\rangle. \tag{8}$$

The time-dependent Hamiltonian includes the field-interaction in length gauge and employing the dipole approximation:

$$H(t) = \frac{p^2}{2} + xE(t) + V(x). \tag{9}$$

The wave function is represented on a spatial grid of [-480, 480] Å with 8192 points. The direct propagation is performed iteratively using the short time propagator with a time step of $\Delta t = 5$ as and applying the split operator technique [33] utilizing the FFTW3 library [34] for Fourier transformation.

The relaxation method [35] is used to obtain the set of bound eigenstates, $\{|\phi_n\rangle\}$, that solve the field-free stationary electronic Schrödinger equation with the time-independent Hamiltonian,

$$H_0 = \frac{p^2}{2} + V(x). \tag{10}$$

In order to suppress reflections at the grid boundaries, we apply the splitting function [36]

$$f(x) = \left[1 + e^{\zeta_1(|x|-\zeta_2)}\right]^{-1} \tag{11}$$

with the parameters $\zeta_1 = 0.19$ Å$^{-1}$ and $\zeta_2 = 480$ Å in every time step.

Contrary to the trajectory-based methods introduced above, in the full quantum dynamical treatment, we cannot easily choose a time interval of birth to be exclusively considered. We will always have contributions overlap and interference of other events, e.g., direct electrons, (bound) polarization, excitation processes, saturation, etc. To minimize those interferences, we chose two different approaches mimicking one and a half cycles of the cw electric field of Eq. 3 of the trajectory-based methods. First, we use a few-cycle pulse defined as

$$E(t) = E_0 e^{-\beta(t-T_0)^2} \cos(\omega(t-T_0) + \phi), \tag{12}$$

using the same parameters as in the cw, Eq. (3), together with a carrier-envelope phase of $\phi = \pi$ and the width of the Gaussian envelope $\beta = \frac{4 \log(2)}{\tau^2}$ determined by the full-width half-maximum of $\tau = 22$ fs. The laser pulse is centered at $T_0 \approx 0.5\, T$, yielding a positive half-cycle at around $t = 0$. Continuum electrons that are born via OBI during this half-cycle return in the following optical cycle and are analyzed using phase-space representation (see below and results). The second approach to the electric field relies on an artificial pulse cut-out of the cw electric field (Eq. (3)). Both electric fields are shown in Figure 7 in Appendix A. These approaches will still lead to competing processes and are discussed in the text. The TDSE is only solved for the angular frequencies of $\omega = 0.014$ a.u. (i.e., $\lambda = 3.2$ μm).

### c. *Phase-space distribution functions*

For comparison and analysis, we calculate (quasi) phase-space probability functions, namely the Wigner [37] and Husimi [38] distributions. This allows to directly compare with classical results [39,40].

For a one-dimensional wave function, $|\Psi(t)\rangle$, the Wigner distribution reads:

$$W(x,p,t) = \frac{1}{\pi} \int dy\, \Psi^*\left(x + \frac{y}{2}, t\right) \Psi\left(x - \frac{y}{2}, t\right) e^{-ipy}. \tag{13}$$

While the Wigner distribution is real-valued, it is not positive definite, making it difficult to interpret in terms of classical phase-space distributions. However, the correct width in coordinate and momentum space are retained yielding the marginals $|\Psi(x)|^2$ or $|\Psi(p)|^2$ by integrating over the complementary variable $p$ or $x$, respectively. The Husimi distribution for a one-dimensional wave function, $|\Psi(t)\rangle$, is given by

$$H(x,p,t) = \frac{1}{2\pi} \left| \int dy\, \Psi(y,t) e^{-\frac{(x-y)^2}{4\sigma^2}} e^{-ipy} \right|^2. \tag{14}$$

It is real and positive definite and can be interpreted as a Gaussian smoothening of the WDF or a Fourier transformation along a position-dependent window function. The parameter $\sigma$ sets the width of the Gaussian convolution in coordinate space and, thus, determines the width in momentum space. It is chosen arbitrarily and, therefore, makes the position and momentum width in HDF arbitrary. However, the choice of $\sigma$ is crucial since depending on the width in position space, structures are averaged over, or depending on the width in momentum space, high-momentum components are added. In this paper, we use a value of $\sigma = 4$ a.u., which was checked against others and ensures that the electronic wave packet is best characterized in position and momentum.

# III. Results

Using the methods previously outlined in the computational details, we obtain properties of *returning* electrons in the OBI regime. First, we revisit the solutions from the SM, compare them to the CT and introduce challenges for the LIED structural retrieval in the OBI regime. Second, we outline a theoretical protocol to obtain *return* times and energies in a fully quantum dynamical approach to validate the trajectory-based methods and their conclusion for the OBI regime in LIED. Thirdly, the problem of short trajectories arising in OBI for LIED is addressed and quantified by simulating the Fourier transform LIED (FT-LIED) analysis [18]. Finally, margins of error introduced by approximating the return properties with the SM for experimental LIED analysis are quantified.

### a. Trajectory-based methods

In Figure 2, we present the results for the trajectory-based methods, SM (2a, c) and CT (2b, d), at 3.2 μm. In particular, we show the distribution of the *return* energy (i.e., laser polarization frame), $E_{ret}$, and *rescattering* energy (i.e., laboratory frame photoelectron data), $E_{resc}$, as a function of the *return* time, $t_{ret}$ for both models (see Eq. (7)).

In the TI regime and standard LIED experiments, one expects long trajectories (i.e., $t_{ret, long} > t_{ret}$ ($E_{ret, max}$)) to contribute most significantly since they are born closer to the electric field peak, which results in an exponentially higher ionization probability compared to the short ones (see Eq. (6)) [15]. However, here in the OBI regime, we observe a significant contribution of short trajectories (i.e., $t_{ret,short} < t_{ret}$ ($E_{ret, max}$)) in the high-energy part of the spectrum as a direct consequence of the overall higher field strength. These contribute also at later times in the cycle, translating into a higher ionization probability originating from short trajectories (see Figure 2). The presence of both short and long trajectories is problematic in LIED because it obscures the one-to-one correspondence between the final *rescattering* energy (i.e., laboratory-frame photoelectron data), $E_{resc,}$ and the energy at the time of *return* (i.e., laser polarization frame), $E_{ret}$, as mathematically described in Eq. (7). This correspondence problem is visualized in more detail in Figure 2: In the experiment, we only measure the rescattering energy, $E_{resc,}$ (panel c,d) but for the structural retrieval, we are interested in the information in the polarization frame (panel a, b) to calculate the momentum transfer and subsequently retrieve structural parameters. The relation of the two frames is Eq. (7): The measured *rescattering* energy contains an additional contribution by the laser field of the magnitude of the vector potential at time of *return*. However, the value of the vector potential varies during the laser cycle, imparting different momenta at the different times of *return*. If only long trajectories are present (as it is for TI), there is a clear one-to-one correspondence regarding which value of the vector potential is added to the *rescattering* energy and, thus, the two frames can be converted trivially. However, in OBI we have additional short trajectories that are born later in the optical cycle and return earlier. These can have the same *rescattering* energy as a long trajectory. Therefore, electrons that arrive to the detector with the same $E_{resc}$ have two different solutions for $E_{ret}$. In Figure 2, the red dashed lines show in panel a and b how a short and long trajectory returning at different times gets imparted different momenta by the vector potential, which leads to the same *rescattering* energy in panel c and d.

The role of the short trajectories in the structural retrieval is analyzed in detail in Section IIIc). The results of SM and CT agree well with each other showing that the influence of the Coulomb potential in the OBI regime is weak. Short trajectories are slightly more pronounced in CT and slightly higher return energies can be reached. Moreover, in the CT solutions, we can observe few additional trajectories returning at later times ($t_{ret} > 1$ opt. cycles). These trajectories are born before the peak of the electric field and are attracted back to the parent ion by the Coulomb potential. These are not LIED relevant trajectories, as they belong to the low-energy part of the spectrum. We will revisit these differences in the final section to quantify the error introduced in the experimental LIED structural retrieval in the OBI regime. We do not consider possible later or multiple returns.

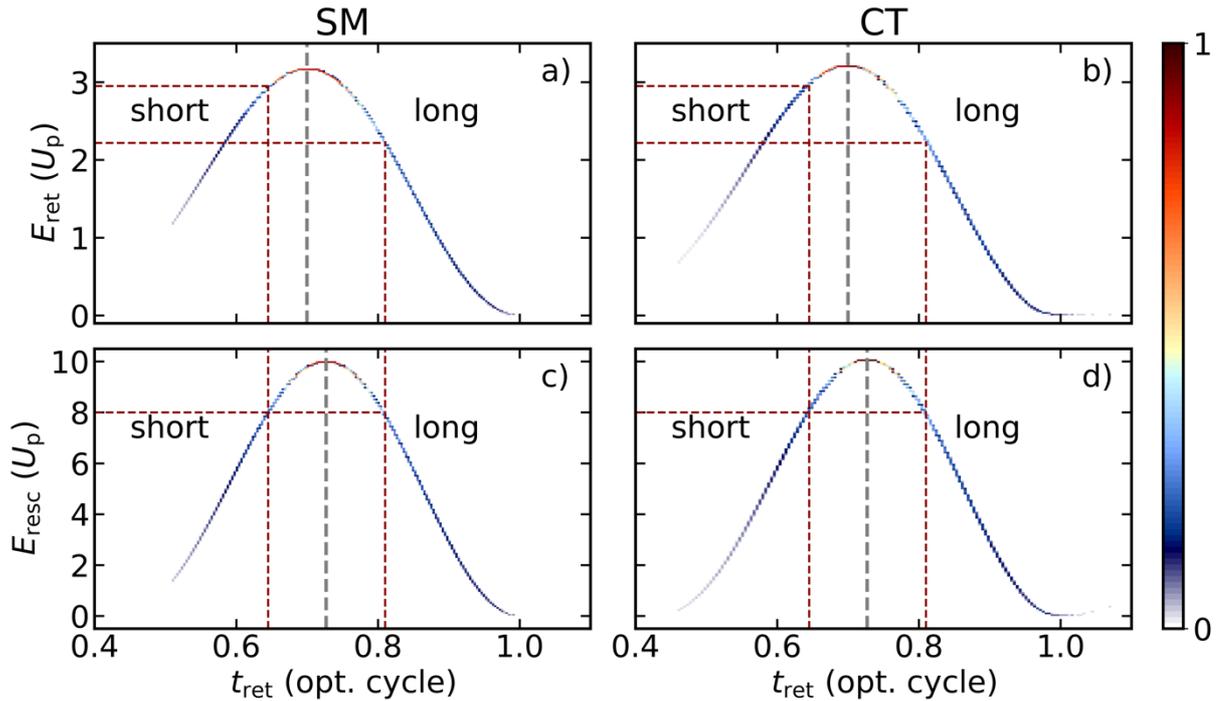

**Figure 2. Correlation histograms showing (a, b) electron *returning* energies, $E_{ret}$, and *rescattering* energies, $E_{resc}$, (c, d) at the time of *return*, $t_{ret}$, obtained with (a, c) the SM and (b, d) the CT for the OBI regime at 3.2 μm.** The color scale mapping represents the normalized intensity as a function of ionization rate. The vertical gray dashed line distinguishes short (left) and long (right) trajectories. The red dashed lines visualize the one-to-one correspondence problem for LIED in the OBI regime and is explained in the text.

### b. *Numerical solution of TDSE*

Next, we compare the trajectory-based methods, SM and CT, to quantum dynamical calculations. As outlined in the computational details, the inherent advantage - but at the same time disadvantage - of the TDSE is that it gives us a complete picture of all processes. To minimize the parallel processes to the *rescattering* part, we firstly use a few-cycle and artificial pulse (recall II b). Secondly, we evaluate the *returning* electrons using the Husimi phase-space representation at a position cut of $x = -20$ Å. Placing the evaluation window away from the nucleus and only regarding momenta in the direction of the nucleus reduces competing contributions from bound electron parts. Moreover, the smoothing effect of the Husimi function averages out additional effects due to interference with previous ionization events. This approach allows us to provide (correlated) *return* energies and *return* times and, subsequently, to compare to the histograms in the trajectory-based methods (Figure 2). The Husimi momentum peak at $x = -20$ Å yields the *return* momentum as the position of the maximum of the peaks (see Figure 3, bottom panels), while the peak intensity allows for quantifying the electron counts of a specific *return* energy (Figure 3, top panels) with respect to the *return* time. A comparison with the SM and CT at $x = -20$ Å is shown as a dark red dotted line and a blue dashed line, respectively, in the bottom panels and as *return* time histograms in the top panels.

The *return* momenta in the bottom panels match very well between all three levels of theory for both the few-cycle and the artificial pulses and validate the trajectory-based approaches. In general, the TDSE *return* energy seems to be slightly lower than the SM and CT results, likely, as an effect of competing processes and especially included bound state effects. The distribution of the *return* times in the top panels matches only partially for all three levels of theory. Perfect coincidence is not expected since the short pulse and artificial pulse constitute an approximation to the cw field in the trajectory-based methods. For the short pulse, the half-cycles are not symmetric and allow for higher momenta to return (see appendix A). Moreover, competing processes, excitation, and saturation challenge the retrieval of *return* properties in TDSE, and the Husimi distribution approach has an inherent ambiguity in terms of the signal width and intensity. Note that during the light pulse interaction we exhibit full depletion of the ground state. Nonetheless, the *return* momenta show excellent agreement and support the results of the trajectory-based methods, showing that they are also valid in such extreme regimes of full depletion.

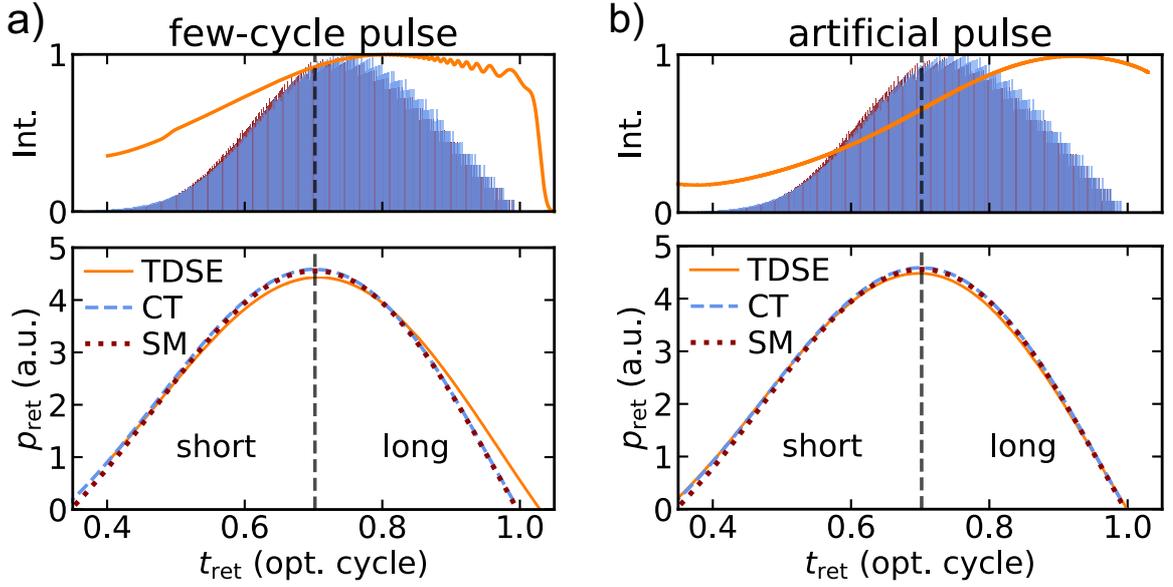

**Figure 3. TDSE phase space results for *return* energies and *return* time using a) a few-cycle and b) an artificial laser pulse of λ = 3.2 μm.** Each bottom panel shows the *return* time, $t_{\mathrm{ret}}$, and *return* momentum, $p_{\mathrm{ret}}$, correlation given by the TDSE (solid orange lines) together with a SM (dotted red lines) and CT (dashed blue lines) result comparison at $x = -20$ Å. Each top panel quantifies the *return* energies of a given *return* time via the phase space intensities. Moreover, the *return* time histograms for the SM (dark red) and CT (blue) are shown. The numerical protocol to obtain this data is outlined in detail in the text.

Additionally, we use the quantum-mechanical phase-space analysis of the TDSE to successfully validate the initial conditions employed in the CT (see Appendix B).

### c. *Fourier transform LIED (FT-LIED) simulation*

We have seen in the previous sections that the SM correctly describes the electron recollision process in the OBI regime, as confirmed by comparing it to the CT and TDSE. However, in all levels of theory, we observe that with the increase of field strength, short trajectories become non-negligible. These trajectories are typically not considered in the structural retrieval methods based on recollision imaging. To this end, we investigate their influence by simulating the entire Fourier transform LIED (FT-LIED) analysis [18,20–22]. The FT-LIED method is based on analyzing the momentum transfer due to scattering only along the laser polarization direction. The FT of this molecular interference signal directly provides the radial distribution of internuclear distances.

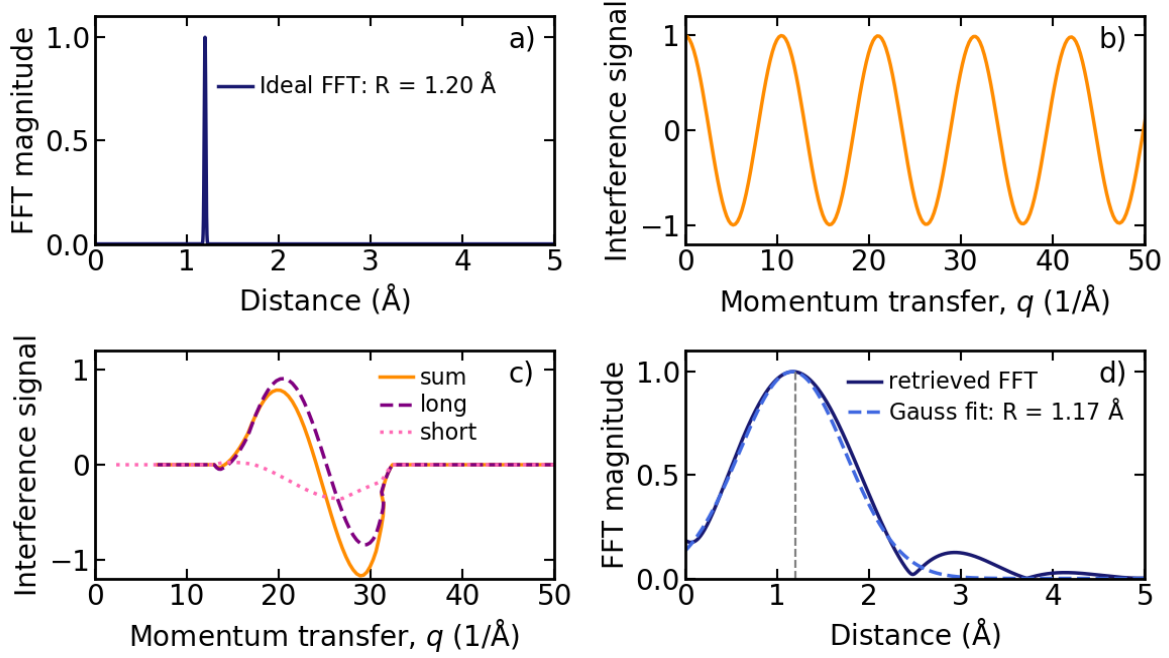

**Figure 4. Description of the FT-LIED simulation** (a) Ideal Fast Fourier spectrum (dark blue) for an arbitrary internuclear distance of a diatomic molecule of $R$ = 1.20 Å, approximated as a Gaussian of width = 0.0005 a.u. (b) Ideal interference signal (orange) in momentum space obtained by IFFT of a). (c) Final interference signal (orange) as a function of momentum transfer, $q = 2p_r$, in the laboratory frame. The signal is cut at the 2-10 $U_p$ rescattering window, for $U_p = E_0^2/(2\omega)^2$ = 94 eV, corresponding to the $E_0$ = 0.053 a.u. and $\omega$ = 0.014 a.u. (i.e., $\lambda$ = **3.2 μm**). A rectangular window and zero padding are applied. d) FFT spectrum (dark blue) along with its gaussian fit (dashed light blue). The vertical grey dashed line represents the initial equilibrium distance of 1.20 Å.

We first generate an electron spectrum for a model diatomic molecule of $I_p$ = 10 eV and an internuclear distance $R$ = 1.20 Å (Figure 4a). This spectrum is approximated by a narrowed Gaussian function with width, $\sigma$ = 5×10$^{-4}$ a.u. By means of the inverse Fast Fourier Transform (IFFT) algorithm [41], we obtain the interference signal in momentum space (Figure 4b) from the spectrum. Next, we cut this infinite interference signal to the corresponding 2-10 $U_p$ *rescattering* window, as we would measure in the laboratory, being $U_p = E_0^2/(2\omega)^2$ = 94 eV for $E_0$ = 0.053 a.u. and $\omega$ = 0.014 a.u. (i.e., $\lambda$ = 3.2 μm). As abovementioned, there exist two possible solutions, corresponding to long and short trajectories (see Figure 2). Each resultant interference signal obtained either for short (dotted pink line in Figure 4c) or long trajectories (dashed purple lines in Figure 3c), is weighted by the corresponding ionization rate (Eq. 6). The overall final interference signal is, therefore, a weighted contribution from the short and long solutions (solid orange line in Figure 4c). This overall signal represents a signal close to that we measure in the lab. Now, to extract the field-free signal, we transform to the laser polarization frame by subtracting the vector potential from the long trajectories' solution, $A_{long}(t_{ret})$, i.e. the vector potential value corresponding to long trajectories only; thus, intentionally neglecting the short trajectory contribution. We follow established experimental procedure by applying a smoothing function [42], zero padding [43] and a rectangular Kaiser window [44]. Next, the interference signal is Fourier transformed to obtain the convoluted FFT (Figure 4d). Finally, we fit a Gaussian (dashed light blue line) to the signal and obtain the retrieved bond length, $R_{ret}$, from its center. We find that the retrieved bond length, $R_{ret}$, exhibits a percentage error (percentage error = (($R_{ret} - R$) / $R$) x 100) of only 2.7% compared to the initially defined bond length $R$. Hence, despite ignoring the increasing contribution of short trajectories, the field-free signal is retrieved with a good accuracy.

We further investigate the effect of short trajectories for a set of different internuclear distances $R$, spanning from 0.75 Å to 2.00 Å, and different wavelengths: 2.0, 3.2 and 4.0 μm. These results are summarized in Figure 5. In general, and as expected, higher $U_p$ values, given by longer wavelengths, in the OBI regime result in a higher accuracy of $R_{ret}$. The better resolution can be explained because a higher *returning* energy translates into a larger momentum transfer window. OBI, however results in an increasing contribution of short trajectories (pink dotted line in Figure 4c) mixing with the total interference signal (solid orange line in Figure 4c). This contribution may add constructively or destructively (purple dashed line in Figure 4c), thus increasing the error in the retrieved bond length up to 12% for realistic experimental parameters. In either case, we find that higher FT sampling in OBI surpasses the short trajectories' contribution to the signal. The overall error is reduced by up to 50% for longer wavelengths. However, one has to keep in mind that for 4 μm in the OBI regime, non-dipole effects become non-negligible (see Table 1, $\beta_0 \geq 1$). The tradeoff between these different constraints make 3.2 μm a sensible choice of wavelength. We find from our investigation that the error due to short trajectory contributions in our FT-LIED structure retrieval methodology is almost negligible, thus validating the one-to-one correspondence between the laboratory frame and the laser polarization frame even for the OBI regime.

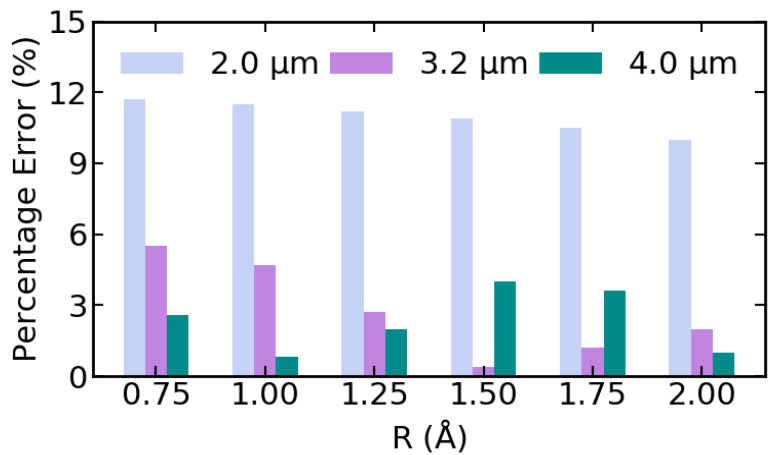

Figure 5. Percentage error of the retrieved internuclear distance, $R_{ret}$, with respect to the initial given $R$ for different wavelengths: 2.0 μm (blue), 3.2 μm (violet) and 4.0 μm

(blue-green) in the OBI regime, obtained with the FT LIED simulated model when ignoring the contribution from short trajectories.

### *d. Classical recollision model: error margin quantification*

Having confirmed that the short trajectories do not significantly alter the LIED structural retrieval in the OBI regime, we now quantify the error margins introduced by the commonly employed SM (neglecting the Coulomb potential). This is achieved by comparing the SM (solid red) to the CT results in the laboratory frame (i.e., 2-10 $U_\text{p}$ scale) for the OBI regimes at three different wavelengths: 2.0 µm (dotted blue), 3.2 µm (dashed violet) and 4.0 µm (dashed-dotted blue-green). Figure 6a) displays the dimensionless ratio of electron *returning* momenta over vector potential at the time of *return*, $p_\text{ret}/|A_\text{ret}|$, at the different *rescattering* energies, $E_\text{resc}$, and Figure 6b), the percentage difference of the CT results with respect to the SM solutions. The vertical gray dashed lines show the 2 $U_\text{p}$ and 10 $U_\text{p}$ cut-offs, defining the *rescattering* window. Since we have shown in the previous section that short trajectories only inflect a small error (e.g., < 5% for 3.2 µm) in the structure retrieval and thus, can be neglected in the structure retrieval in the OBI regime, we choose only solutions from long trajectories. For the three different wavelengths employed in the present work, the percentage difference of the CT with respect to the SM is below 2% within the *rescattering* plateau (i.e., 2-10 $U_\text{p}$), showing an excellent agreement. In the MIR, this error is further reduced to an entirely negligible ~0.5%, indicating a clear advantage of MIR wavelengths, as long as the dipole approximation remains valid. Again, this demonstrates why 3.2 µm wavelength is our optimal choice for LIED.

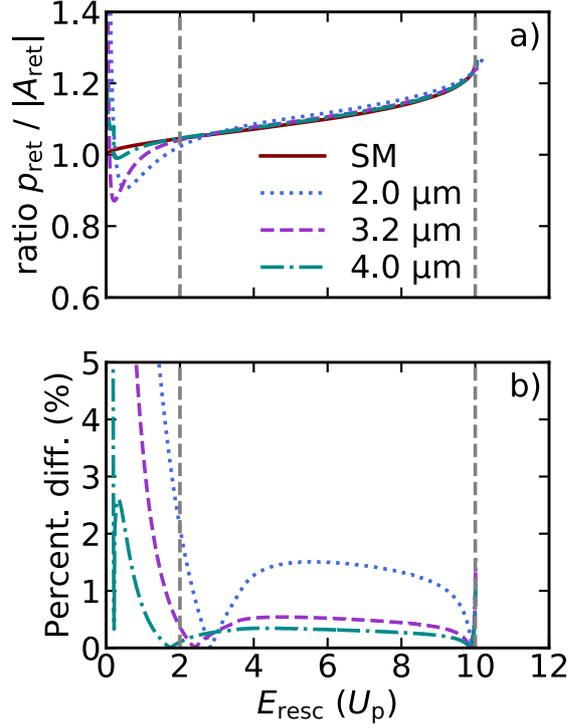

**Figure 6. Quantification of the error margin of the SM (solid dark-red lines) with respect to the CT as a function of photoelectron energies after *rescattering* $E_{\text{resc}}$, for the OBI regime and for three different wavelengths: 2.0 μm (dotted blue), 3.2 μm (dashed violet) and 4.0 μm (dashed-dotted blue-green).** a) The ratio of electron *returning* momenta over vector potential at the time of *return*, $p_{\text{ret}}/|A_{\text{ret}}|$, at the different *rescattering* energies, $E_{\text{resc}}$. b) The percentage difference of the CT results with respect to the SM solutions. The grey vertical dashed lines show the $2U_p$ and $10U_p$ classical cut-offs of direct and *rescattered* electrons, respectively, which is the range normally employed to evaluate LIED results.

## IV. Conclusion

Recollision imaging is a powerful attosecond technology, which is easy to implement, but whose information extraction relies on unambiguous momentum-to-time mapping by invoking the three-step model, also called the simple-man's model, of strong-field physics. This model works very well in the quasi-static, or tunneling, regime and, amongst other simplifications, it assumes that the recolliding (imaging) electron is born with zero initial kinetic energy. However, recollision imaging of larger molecular structures requires high peak intensity for recollision to occur despite the much lower ionization energy of the target. Consequently, recollision occurs in the over-the-barrier regime for which initial kinetic energies are non-zero and short trajectories contributions are significant. We conducted a numerical study of FT-LIED, based on classical trajectory dynamics, the simple man's model, and the full quantum dynamics by solving the TDSE, and show that the accrued intrinsic error is still very small. For realistic assumptions, we find structural retrieval errors below 5% (> 3.2 μm driving laser), thus justify the usage of the SM despite the OBI regime. Considering the long trajectories solely, the error margin of the SM with respect to the CT is less than 2% for realistic wavelengths and less than 0.5% for the MIR. This study provides a foundation for extending recollision-based imaging methods to low-$I_p$ atomic, molecular or liquid systems.

## Acknowledgements

J.B. acknowledges financial support from the European Research Council for ERC Advanced Grant "TRANSFORMER" (788218), ERC Proof of Concept Grant "miniX" (840010), FET-OPEN "PETACom" (829153), FET-OPEN "OPTOlogic" (899794), Laserlab-Europe (654148), Marie Sklodowska-Curie ITN "smart-X" (860553), MINECO for Plan Nacional FIS2017-89536-P; AGAUR for 2017 SGR 1639, MINECO for "Severo Ochoa" (SEV- 2015-0522), Fundació Cellex Barcelona, the CERCA Programme / Generalitat de Catalunya, and the Alexander von Humboldt Foundation for the Friedrich Wilhelm Bessel Prize. J.B. and A.S. acknowledge funding from the Marie Sklodowska-Curie grant agreement No. 641272. J.B and B.B. acknowledge funding (PRE2019-088522) from MCIN/AEI/10.13039/501100011033 y FSE "El FSE invierte en tu futuro". S.G. acknowledges support from the European Research Council (ERC) for the ERC Consolidator Grant QUEM-CHEM (772676). K.M.Z. and S.G. are part of the Max Planck School of Photonics supported by BMBF, Max Planck Society, and Fraunhofer Society.

# APPENDIX A: ADDITIONAL COMPUTATIONAL DETAILS

The two different approaches to mimic the cw electric field employed in the trajectory-based methods are shown in Figure 7. The few-cycle pulse as described in Eq. (12) is shown as a solid blue line and the artificial pulse cut-out of the cw electric field from Eq. (3) is represented on top with dashed red line.

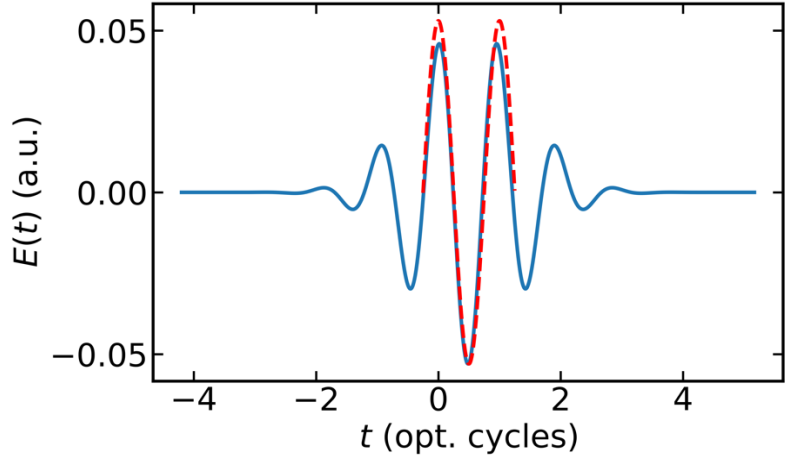

**Figure 7. Electric fields used in the numerical solution of the TDSE at 3.2 μm.** The few-cycle pulse as described in Eq. (12) (solid blue line) together with the artificial pulse (dashed red line) as cut-out of continuous wave (cw) electric field, Eq. (3).

The *funnel effect* and the trigonometric fit are exemplary visualized for a trajectory started at $t_0 = 0.05$ opt. cycles in Figure 8.

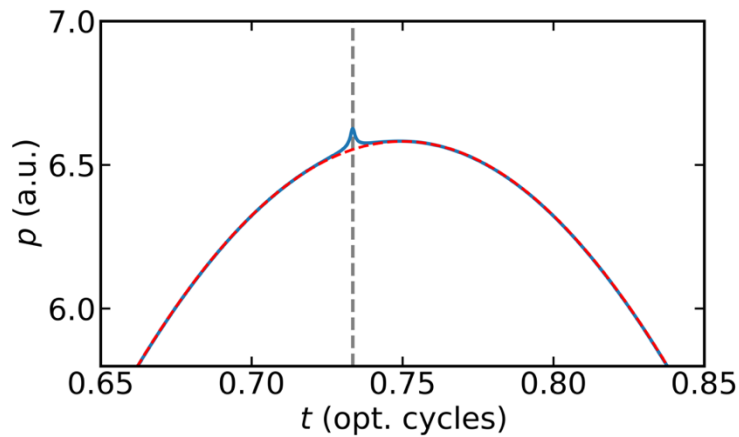

**Figure 8. Momentum behavior of a selected trajectory started at $t_0$ = 0.05 opt. cycles.** The red dashed line represents the trigonometric fit to the momenta, the vertical gray dashed line indicates the *return* time.

# APPENDIX B: Validation of the initial momenta in the CT via TDSE

As outlined in the text, we analyze the phase-space representation of the TDSE via the Wigner and Husimi function to validate the initial momenta in the CT for the OBI regime. An example is shown in Figure 9. The upper panel shows a colormap of the Wigner phase-space together with a yellow horizontal line indicating the CT's initial position, $x_0^{OBI}(t_0)$, at time of peak electric field strength, $t_0$. The lower panel shows the momentum distributions at this very position, $x_0^{OBI}(t_0)$ given by cut-along the phase-space distributions of the Wigner and Husimi representation. Contrary to the TI regime, where typically the trajectories are assumed to start with $p_0^{TI}(t_0) = 0$, in OBI the electrons are born with $p_0^{OBI}(t_0)$ different from zero. The TDSE results show that the classical initial momentum is in good agreement with the Wigner representation, see Figure 7. The Husimi representation cut is at slightly lower momenta. This is due to the fact that while the Wigner representation gives the momentum distribution along the exact cut, (i.e., the CT initial position $x_0^{OBI}(t_0)$), the Husimi representation yields the momentum along a Gaussian coordinate window, which is centered at the CT initial position $x_0^{OBI}(t_0)$. Hence, it incorporates the momentum distribution around $x_0^{OBI}(t_0)$, which includes low momentum parts closer to the nucleus. This shifts the overall Husimi momentum cut to lower momenta. Nonetheless, the classical approximation is in good agreement with the TDSE phase space results. Furthermore, the Wigner momentum distribution at the initial position shows that most part of the momentum distribution is negative, i.e., pointing in the direction of the lowered dressed Coulomb potential.

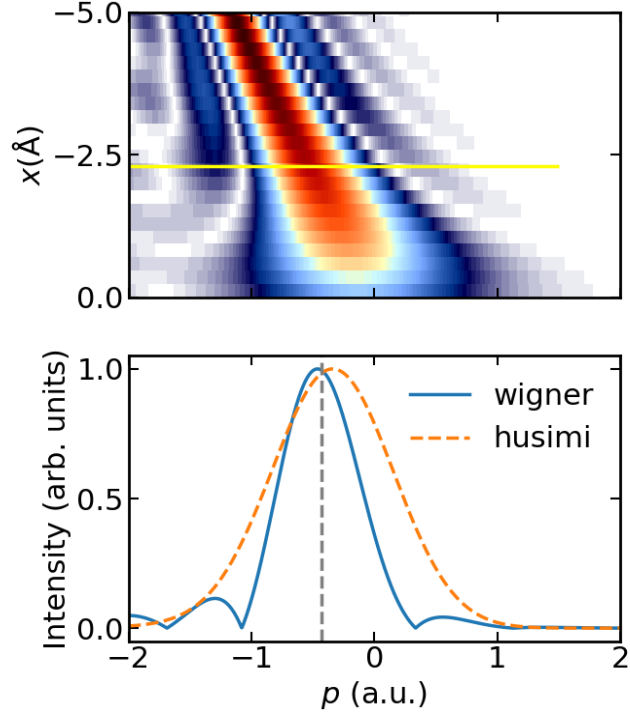

**Figure 9. Wigner and Husimi phase space representation snapshot of the TDSE at peak intensity.** The peak intensity corresponds $E(t_0) = E_0 = 0.053$ a.u. The top panel shows the colormap of the absolute Wigner representation with the yellow horizontal line indicating the classical CT initial exit position, $x_0^{\text{OBI}}(t_0)$ (Eq. (4)). The lower panel shows the momentum distribution cut in Wigner and Husimi representation along the classical CT initial exit position, i.e., the yellow line in the top panel. The vertical gray dashed line represents the CT initial momenta, $p_0^{\text{OBI}}(t_0)$.

# References

[1] B. Wolter, M. G. Pullen, M. Baudisch, M. Sclafani, M. Hemmer, A. Senftleben, C. D. Schröter, J. Ullrich, R. Moshammer, and J. Biegert, Strong-Field Physics with Mid-IR Fields, Phys Rev X **5**, 021034 (2015).

[2] M. Nisoli, P. Decleva, F. Calegari, A. Palacios, and F. Martín, Attosecond Electron Dynamics in Molecules, Chem Rev **117**, 10760 (2017).

[3] J. Xu, C. I. Blaga, P. Agostini, and L. F. DiMauro, Time-Resolved Molecular Imaging, Journal of Physics B: Atomic, Molecular and Optical Physics **49**, 112001 (2016).

[4] H. R. Reiss, Limits on Tunneling Theories of Strong-Field Ionization, Phys Rev Lett **101**, 043002 (2008).

[5] H. R. Reiss, Novel Phenomena in Very-Low-Frequency Strong Fields, Phys Rev Lett **102**, 143003 (2009).

[6] C. Rischel, A. Rousse, I. Uschmann, P.-A. Albouy, J.-P. Geindre, P. Audebert, J.-C. Gauthier, E. Fröster, J.-L. Martin, and A. Antonetti, Femtosecond Time-Resolved X-Ray Diffraction from Laser-Heated Organic Films, Nature **390**, 490 (1997).

[7] P. B. Corkum, Plasma Perspective on Strong Field Multiphoton Ionization, Phys Rev Lett **71**, 1994 (1993).

[8] F. H. M. Faisal, Strong-Field Physics: Ionization Surprise, Nat Phys **5**, 319 (2009).

[9] W. Becker, F. Grasbon, R. Kopold, D. B. Milošević, G. G. Paulus, and H. Walther, Above-Threshold Ionization: From Classical Features to Quantum Effects, Advances in Atomic, Molecular and Optical Physics **48**, 35 (2002).

[10] W. Quan et al., Classical Aspects in Above-Threshold Ionization with a Midinfrared Strong Laser Field, Phys Rev Lett **103**, 093001 (2009).

[11] C. I. Blaga, F. Catoire, P. Colosimo, G. G. Paulus, H. G. Muller, P. Agostini, and L. F. DiMauro, Strong-Field Photoionization Revisited, Nat Phys **5**, 335 (2009).

[12] M. Lewenstein, Ph. Balcou, M. Yu. Ivanov, A. L'Huillier, and P. B. Corkum, Theory of High-Harmonic Generation by Low-Frequency Laser Fields, Phys Rev A **49**, 2117 (1994).

[13] B. Walker, B. Sheehy, L. F. DiMauro, P. Agostini, K. J. Schafer, and K. C. Kulander, Precision Measurement of Strong Field Double Ionization of Helium, Phys Rev Lett **73**, 1227 (1994).

[14] T. Zuo, A. D. Bandrauk, and P. B. Corkum, Laser-Induced Electron Diffraction: A New Tool for Probing Ultrafast Molecular Dynamics, Chem Phys Lett **259**, 313 (1996).

[15] M. Meckel et al., Laser-Induced Electron Tunneling and Diffraction, Science (1979) **320**, 1478 (2008).

[16] J. Xu, Z. Chen, A. T. Le, and C. D. Lin, Self-Imaging of Molecules from Diffraction Spectra by Laser-Induced Rescattering Electrons, Phys Rev A **82**, 33403 (2010).

[17] C. Yu, H. Wei, X. Wang, A. T. Le, R. Lu, and C. D. Lin, Reconstruction of Two-Dimensional Molecular Structure with Laser-Induced Electron Diffraction from Laser-Aligned Polyatomic Molecules, Sci Rep **5**, 15753 (2015).

[18] M. G. Pullen et al., Influence of Orbital Symmetry on Diffraction Imaging with Rescattering Electron Wave Packets, Nat Commun **7**, 11922 (2016).

[19] N. B. Delone and V. P. Krainov, Tunneling and Barrier-Suppression Ionization of Atoms and Ions in a Laser Radiation Field, Physics-Uspekhi **41**, 469 (1998).

[20] B. Belsa et al., Laser-Induced Electron Diffraction of the Ultrafast Umbrella Motion in Ammonia, Structural Dynamics **8**, 014301 (2021).

[21] J. Xu, C. I. Blaga, K. Zhang, Y. H. Lai, C. D. Lin, T. A. Miller, P. Agostini, and L. F. DiMauro, Diffraction Using Laser-Driven Broadband Electron Wave Packets, Nat Commun **5**, 4635 (2014).

[22] X. Liu et al., Imaging an Isolated Water Molecule Using a Single Electron Wave Packet, Journal of Chemical Physics **151**, 024306 (2019).

[23] C. I. Blaga, J. Xu, A. D. DiChiara, E. Sistrunk, K. Zhang, P. Agostini, T. A. Miller, L. F. DiMauro, and C. D. Lin, Imaging Ultrafast Molecular Dynamics with Laser-Induced Electron Diffraction, Nature **483**, 194 (2012).

[24] B. Wolter et al., Ultrafast Electron Diffraction Imaging of Bond Breaking in Di-Ionized Acetylene, Science (1979) **354**, 308 (2016).

[25] K. Amini et al., Imaging the Renner–Teller Effect Using Laser-Induced Electron Diffraction, Proc Natl Acad Sci U S A **116**, 8173 (2019).

[26] A. Sanchez et al., Molecular Structure Retrieval Directly from Laboratory-Frame Photoelectron Spectra in Laser-Induced Electron Diffraction, Nat Commun **12**, 1520 (2021).

[27] H. Fuest et al., Diffractive Imaging of C60 Structural Deformations Induced by Intense Femtosecond Midinfrared Laser Fields, Phys Rev Lett **122**, 053002 (2019).

[28] Y. Ito, C. Wang, A. T. Le, M. Okunishi, D. Ding, C. D. Lin, and K. Ueda, Extracting Conformational Structure Information of Benzene Molecules via Laser-Induced Electron Diffraction, Structural Dynamics **3**, 034303 (2016).

[29] Y. Ito, R. Carranza, M. Okunishi, R. R. Lucchese, and K. Ueda, Extraction of Geometrical Structure of Ethylene Molecules by Laser-Induced Electron Diffraction Combined with Ab Initio Scattering Calculations, Phys Rev A **96**, 053414 (2017).

[30] E. T. Karamatskos et al., Atomic-Resolution Imaging of Carbonyl Sulfide by Laser-Induced Electron Diffraction, J Chem Phys **150**, 244301 (2019).

[31] M. v Ammosov, N. B. Delone, and V. P. Krainov, Tunnel Ionization of Complex Atoms and of Atomic Ions in an Alternating Electromagnetic Field, Sov. Phys. JETP **64**, 1191 (1986).

[32] X. M. Tong and C. D. Lin, Empirical Formula for Static Field Ionization Rates of Atoms and Molecules by Lasers in the Barrier-Suppression Regime, Journal of Physics B: Atomic, Molecular and Optical Physics **38**, 2593 (2005).

[33] M. D. Feit, J. A. Fleck Jr, and A. Steiger, Solution of the Schrödinger Equation by a Spectral Method, J Comput Phys **47**, 412 (1982).

[34] M. Frigo and S. G. Johnson, FFTW: An Adaptive Software Architecture for the FFT, in Proceedings of the 1998 IEEE International Conference on Acoustics, Speech and Signal Processing, ICASSP'98 (Cat. No. 98CH36181), Vol. 3 (IEEE, 1998), pp. 1381–1384.

[35] R. Kosloff and H. Tal-Ezer, A Direct Relaxation Method for Calculating Eigenfunctions and Eigenvalues of the Schrödinger Equation on a Grid, Chem Phys Lett **127**, 223 (1986).

[36] R. Heather and H. Metiu, An Efficient Procedure for Calculating the Evolution of the Wave Function by Fast Fourier Transform Methods for Systems with Spatially Extended Wave Function and Localized Potential, J Chem Phys **86**, 5009 (1987).

[37] E. P. Wigner, On the Quantum Correction for Thermodynamic Equilibrium, in Part I: Physical Chemistry. Part II: Solid State Physics (Springer, 1997), pp. 110–120.

[38]   K. Husimi, Some Formal Properties of the Density Matrix, Proceedings of the Physico-Mathematical Society of Japan. 3rd Series **22**, 264 (1940).

[39]   S. Gräfe, J. Doose, and J. Burgdörfer, Quantum Phase-Space Analysis of Electronic Rescattering Dynamics in Intense Few-Cycle Laser Fields, Journal of Physics B: Atomic, Molecular and Optical Physics **45**, 55002 (2012).

[40]   M. Hillery, R. F. O'Connell, M. O. Scully, and E. P. Wigner, Distribution Functions in Physics: Fundamentals, Phys Rep **106**, 121 (1984).

[41]   J. W. Cooley and J. W. Tukey, An Algorithm for the Machine Calculation of Complex Fourier Series, n.d.

[42]   L. Hörmander, Convolution, in The Analysis of Linear Partial Differential Operators I, Vol. 256 (1998), pp. 87–125.

[43]   J. O. (Julius O. Smith, Mathematics of the Discrete Fourier Transform (DFT): With Audio Applications, Second Edi (W3K Publishing, 2007).

[44]   K. M. M. Prabhu, Window Functions and Their Applications in Signal Processing (CRC Press, 2018).